\documentclass[11pt]{amsart}
\reversemarginpar
\newcommand{\commentout}[1]{}

\usepackage{amsmath}
\usepackage{amssymb}

\newcommand{\nwc}{\newcommand}

 \relax
\newif\ifatten\attenfalse
 \relax
\newcommand{\atten}[1]{%
  \unskip
    \ifatten
    \raisebox{2pt}{\ \colorbox{yellow}{\hspace{.25cm}}}
      \fi
      }

\newcommand{\lt}{\left}
\newcommand{\rt}{\right}
\newcommand{\vas}{\varepsilon}
\newcommand{\lan}{\left\langle}
\newcommand{\ran}{\right\rangle}
\newcommand{\tvas}{T_t^\vas}
\newcommand{\vvas}{\tilde{V}_t^\vas}
\newcommand{\veptil}{\tilde{V}_t^\vas}
\newcommand{\vep}{{V}_t^\vas}
\newcommand{\cv}{{\ml V}^\ep_t}
\newcommand{\cvtil}{\tilde{{\ml V}}^\ep_t}
\newcommand{\n}{\nabla}

\nwc{\nwt}{\newtheorem}

\nwt{prop}{Proposition}
\nwt{proposition}{Proposition}
\nwt{lemma}{Lemma}
\nwt{theorem}{Theorem}
\nwt{cor}{Corollary}
\nwt{rem}{Remark}
\nwt{remark}{Remark}
\nwt{defin}{Definition} 

\nwc{\bal}{\begin{align}}
\nwc{\be}{\begin{equation}}
\nwc{\ben}{\begin{equation*}}
\nwc{\bea}{\begin{eqnarray}}
\nwc{\beq}{\begin{eqnarray}}
\nwc{\bean}{\begin{eqnarray*}}
\nwc{\beqn}{\begin{eqnarray*}}
\nwc{\beqast}{\begin{eqnarray*}}

\nwc{\eal}{\end{align}}
\nwc{\ee}{\end{equation}}
\nwc{\een}{\end{equation*}}
\nwc{\eea}{\end{eqnarray}}
\nwc{\eeq}{\end{eqnarray}}
\nwc{\eean}{\end{eqnarray*}}
\nwc{\eeqn}{\end{eqnarray*}}
\nwc{\eeqast}{\end{eqnarray*}}

\nwc{\ep}{\varepsilon}
\nwc{\vrho}{\varrho}
\nwc{\orho}{\bar\varrho}
\nwc{\ou}{\bar u}
\nwc{\vpsi}{\varpsi}
\nwc{\lamb}{\lambda_\varepsilon}

\nwc{\nn}{\nonumber}
\nwc{\bm}{\boldmath}
\nwc{\mf}{\mathbf}
\nwc{\ml}{\mathcal}

\nwc{\IA}{\mathbb{A}} 
\nwc{\IB}{\mathbb{B}}
\nwc{\IC}{\mathbb{C}} 
\nwc{\ID}{\mathbb{D}} 
\nwc{\IM}{\mathbb{M}} 
\nwc{\IP}{\mathbb{P}} 
\nwc{\IE}{\mathbb{E}} 
\nwc{\IF}{\mathbb{F}} 
\nwc{\IG}{\mathbb{G}} 
\nwc{\IN}{\mathbb{N}} 
\nwc{\IQ}{\mathbb{Q}} 
\nwc{\IR}{\mathbb{R}} 
\nwc{\IT}{\mathbb{T}} 
\nwc{\IZ}{\mathbb{Z}} 

\nwc{\cE}{{\ml E}}
\nwc{\cV}{{\ml V}}
\nwc{\cC}{{\ml C}}
\nwc{\cA}{{\ml A}}
\nwc{\cL}{{\ml L}}
\nwc{\cK}{{\ml K}}
\nwc{\cB}{{\ml B}}
\nwc{\cD}{{\ml D}}
\nwc{\cF}{{\ml F}}
\nwc{\cM}{{\ml M}}
\nwc{\cG}{{\ml G}}
\nwc{\cH}{{\ml H}}

\begin{document}

\title{Convergence of Passive Scalar Fields in Ornstein-Uhlenbeck Flows
to Kraichnan's Model}

\author{Albert C. Fannjiang}
\thanks{Department of Mathematics,
University of California at Davis,
Davis, CA 95616
Internet: fannjian@math.ucdavis.edu
The research is supported in part by the grant from U.S. National
Science Foundation, DMS-9971322 and the UC Davis Chancellor's Fellowship}

\begin{abstract}
We prove that the passive scalar field in the Ornstein-Uhlenbeck velocity
field with wave-number dependent correlation times converges, in
the white-noise limit, to that of
Kraichnan's model with higher spatial regularity.
\end{abstract}

\maketitle

\section{Introduction}
A passive scalar field $T(t,x)$ in a given fluid velocity  $u(t,x)$
satisfies the advection-diffusion equation
\be
\label{ad}
\frac{\partial T}{\partial t}= u\cdot\nabla T+\frac{\kappa}{2}\Delta T,\quad
T(0,x)=T_0(x)
\ee
where $\kappa \geq 0$ is the molecular diffusivity.
Kraichnan's  model for passive scalar
 has been widely studied to understand turbulent
 transport in the inertial range because of its tractability
(see, e.g.,  
\cite{Ma}, \cite{LR}, \cite{FGV} and the
references therein). 
The model and its variant postulate
a white-noise-in-time, {\em compressible or incompressible}
 velocity field $u$ 
which can be described as
the time derivative of a zero mean, isotropic
Brownian field $B_t$ with the  two-time structure function
\beq
\nonumber
\lefteqn{\IE [B_t(x)-B_t(y)]\otimes[B_s(x)-B_s(y)]}\\
&=&\min{(t,s)}\int 2[\exp{(ik\cdot (x-y))}-1]a^{-1}\cE(\eta+1,k)|k|^{1-d}dk,\quad a>0
\label{4B}
\eeq
 Here
$2a^{-1}\cE(\eta+1,k)$ is the spatial power spectrum with
\[
\cE(\eta+1, k)= E_0(k) |k|^{-2\eta-1}\quad \hbox{for} \,\, \ell_0^{-1}
\ll |k|\ll \ell_1^{-1},\quad \eta\in (0,1)
\]
where $E_0(k)$ is a positive-definite matrix whose
entries are homogeneous functions of degree zero,
$\ell_0$ and $\ell_1$ are the integral and
the viscous scales respectively and they determine the so-called
inertial range. Below the viscous scale $\ell_1$ the velocity field is smooth.
The spatial Hurst exponent $\eta$ characterizes the roughness
of the velocity field in the inertial range and equals $1/3$ in the case of
Kolmogorov's theory of turbulence. 
The tractability of
this model lies in the Gaussian and white-noise nature of
the velocity field.
To fix the idea, we interpret 
eq. (\ref{ad}) in the sense of Stratonovich's integral
\be
\label{kr}
d T = \nabla T\circ d B_t+ \frac{\kappa_0}{2} \Delta T \,\,dt,\quad
\kappa_0\geq 0,\quad T(0,x)=T_0(x).
\ee

To study the effect of a more realistic temporal structure,
one naturally considers the Ornstein-Uhlenbeck (OU) velocity
field 
\be
\label{ad'}
u(t,x)=\frac{1}{\ep} V(\frac{t}{\ep^2}, x)
\ee
with a similar spatial structure but
a wave-number-$k$ dependent correlation time $a^{-1}|k|^{-2\beta},\,\,
a>0, \beta>0$, 
where $\ep>0$ is the scaling parameter.
The two-time structure function has 
the spectral representation
\beq
\label{ou}
&&\IE [V(t,x)-V(t,y)]\otimes[V(s,x)-V(s,y)]\\
&&=\int_{\IR^d} [\exp{(i k\cdot (x-y))}-1]
\exp{(-a|k|^{2\beta}|t-s|)}
\cE(\alpha, k) |k|^{1-d}d k
\nonumber
\eeq
where $\cE$ is the power spectrum given by (\ref{power})
 with $ 1<\alpha < 2$
(see \cite{phd}, \cite{FKP}).
The spatial Hurst exponent of the velocity equals $\alpha-1$
in the inertial range.
The parameters $\alpha, \beta$ have the value $4/3, 1/3$,
respectively, in the case
of Kolmogorov's theory of turbulence.

In this paper we study the relation between these two model.
For simplicity of the presentation we set 
\be
\label{power}
\cE(\alpha, k)=\left\{\begin{array}{ll}
E_0(k)|k|^{1-2\alpha},&\hbox{for}\,\,|k|\in (\ell_0^{-1},
\ell_1^{-1})\\
0,& \hbox{for}\,\, |k|\not\in (\ell_0^{-1},\ell_1^{-1}).
\end{array}
\rt.
\ee
with $\ell_0<\infty, \ell_1>0$.
We defer the discussion of the meaning of solutions of
(\ref{ad}) and (\ref{kr}) until Section~2.

First we consider the situation of 
a non-vanishing ultraviolet cutoff $\ell_1>0$.
We have the following correspondence principle.
\begin{theorem}
Let $\ell_0<\infty, \ell_1>0$ be fixed. Let $\kappa=\kappa(\ep)\geq 0$ 
and $\lim_{\ep\to 0}\kappa=\kappa_0<\infty$. Let $T_0\in L^\infty
(\IR^d)$.

Then 
 the solution $T_t^\ep$ of (\ref{ad}) with the drift
 (\ref{ad'}) converges in distribution,
 as $\ep\to 0$,
  in the space $D([0,t_0);L^\infty_{w^*}(\IR^d)),
  \forall t_0<\infty$
  to the unique
  solution $T_t$ of the martingale problem
   (cf. (\ref{13.2})) corresponding to eq. (\ref{kr}), where the Brownian 
   velocity field
    has the spatial covariance
    with the power spectrum $2a^{-1}\cE(\alpha+\beta,k)$.
	    Here $D([0,t_0);L^\infty_{w^*}(\IR^d))$
    is the space of $L^\infty(\IR^d)$-valued
	      right continuous processes with left limits endowed
       with the Skorohod metric \cite{Bi} and $L^\infty_{w^*}(\IR^d)$
        is
 the standard space $L^\infty(\IR^d)$
		  endowed with the weak* topology.
\end{theorem}
This result suggests that in the limit of rapid temporal decorrelation
 the OU flow
resembles Kraichnan's model with a higher spatial regularity
$\eta=\alpha+\beta-1$.  In particular,  the strict
Kolmogorov's theory $\alpha=4/3,\beta=1/3$ now corresponds
to $\eta=2/3$ in Kraichnan's model.

In the next theorem we let $\ell_1$ vanish along with the scaling
factor $\ep$.
In such a  limit Theorem~1 is not expected to hold for
{\em compressible flows} in the entire range
of $\alpha,\beta$ for the Stratonovich correction term in
the limiting Kraichnan model is well-defined only if 
$\alpha+\beta>3/2$. Moreover, for $\alpha+\beta<2$ and $\ell_1=0$,
the Kraichnan model with {\em compressible}
velocity field may not have a unique solution for a given
initial condition due to the spatial non-Lipschitzness
of the velocity field (cf. \cite{GV}, \cite{LR}).
\begin{theorem}
Suppose that the OU velocity field $V$ is divergence-free, $\n\cdot V=0$.
Let $\ell_0<\infty$  be fixed and $\ell_1=\ell_1(\ep)>0$ such that
$\lim_{\ep\to 0}\ell_1=0$. 
Let $\kappa=\kappa(\ep)\geq 0, \lim_{\ep\to 0}\kappa=\kappa_0<\infty.$
Let $T_0\in L^\infty\cap L^2(\IR^d)$. 
If, additionally,  any one of
the following conditions is satisfied:
\begin{itemize}
\item[(i)] $\alpha+2\beta> 4$;
\item[(ii)] $\alpha+2\beta=4,\quad \lim_{\ep\to 0}\kappa\ep^2
\sqrt{\log{(1/\ell_1)}}=0$;
\item[(iii)] $3<\alpha+2\beta<4,\quad\lim_{\ep\to 0}\kappa\ep^2\ell_1^{\alpha+2\beta-4}=0$;
\item[(iv)] $\alpha+2\beta=3,\quad\lim_{\ep\to 0}\ep
\sqrt{\log{(1/\ell_1)}}=\lim_{\ep\to 0}
\kappa\ep^2\ell_1^{-1}=0$;
\item[(v)] $2<\alpha+2\beta<3,\quad
\lim_{\ep\to 0}\ep\ell_1^{\alpha+2\beta-3}=
\lim_{\ep\to 0}\kappa\ep^2\ell_1^{\alpha+2\beta-4}=0$;
\item[(vi)] $\alpha+2\beta\leq 2,
\quad\lim_{\ep\to 0}\ep\ell_1^{\alpha+2\beta-3}=0$,
\end{itemize}
then the  convergence holds as 
in Theorem~1 but in the space
$D([0,t_0);L^\infty_{w^*}\cap L_w^2(\IR^d)),\forall t_0<\infty$ where
$ L_w^2(\IR^d)$ is the usual $L^2$-function space
endowed with the weak topology.
The Brownian flow of the limiting Kraichnan's model has the
the spatial power spectrum $2a^{-1}\bar{\cE}(\alpha+\beta,k)$
where
\[
\bar{\cE}(\alpha+\beta,k)=\lim_{\ell_1\to 0}\cE(\alpha+\beta,k).
\]
\end{theorem}
\begin{remark}
The assumption of $L^2(\IR^d)$-initial condition in Theorem~2 is to
ensure uniqueness of the limiting Kraichnan model with
$\ell_1=0$ (see Section~2). The limiting velocity field
is only spatially H\"{o}lder continuous (for $\alpha+\beta<2$) with exponent
$\alpha+\beta-1$.
\end{remark}
\begin{remark}
In Theorem~2, when $\kappa_0>0$ and $2<\alpha+2\beta<3$,
$\lim_{\ep\to 0}\kappa\ep^2\ell_1^{\alpha+2\beta-4}=0$ implies
$\lim_{\ep\to 0}\ep\ell_1^{\alpha+2\beta-3}=0$.
\end{remark}
\begin{remark}
In the special case of $\kappa_0=0$, the limiting
Kraichnan model preserves the $L^2$-norm of the initial
condition. On the other hand, the energy identity
for the pre-limiting model
(\cite{LU}, Chapt. III, Theorem 7.2)
\be
\label{45}
\int |\tvas(x)|^2 \,dx+\kappa\int^t_0\int |\n \tvas|^2(x)\,dx\,ds
=\int|T_0(x)|^2\,dx
\ee
implies that $
\|\tvas\|_2 <\|T_0\|_2.
$
Consequently, the convergence in the sense of
the weak-$L^2$ topology in Theorem~2 implies
that $\lim_{\ep\to
0} \|\tvas\|_2 = \|T_t\|_2$ and that the convergence
is indeed in the strong
$L^2$ sense.
\end{remark}

\commentout{
The condition $\alpha+2\beta>2$ is to ensure the spatial differentiability
of the field
\be
\label{6}
\int^\infty_t\IE_t[V(s,\cdot)]\,ds,
\ee
with $\ell_1=0$,
where $\IE_t$ is the expectation conditioning
on the events up to $t$. The field (\ref{6}) has a power spectrum
$a^{-1}\cE(\alpha+2\beta,k)$.
}

Finally we note that the Gaussianity of the velocity field
is not essential to the results. It
has been
used in the proofs to control the first 4 moments of the velocity fields
and to have a mild decay in
the tail distributions of the velocity fields
(cf. (\ref{1.4})). The comparable result in \cite{Kun} requires
a faster-than-Gaussian decay in the tail distributions and
does not apply here. It also requires spatial regularity in the velocity 
fields.

\section{Formulation of solutions}

\commentout{
Convergence of O-U model to Kraichnan's model:
\begin{equation*}
\frac{\partial T^\varepsilon}{\partial t}=\frac{\kappa}{2}\, \Delta
T^\varepsilon + \frac{1}{\varepsilon}V \left(x,
\frac{t}{\varepsilon^2}\right)\cdot \nabla T^\varepsilon, \quad
T^\varepsilon(x,0)=T_0(x)\in C^\infty (\mathbb{R}^d).
\end{equation*}
Let $V(t,x)$ be a O-U flow with the 2-point 2-time correlation
function
\begin{equation*}
\mathbb{E}\{V(t,x)\, V(s,y)\}=\int e^{i(x-y)\cdot
k}\, \varepsilon(k)\, \rho(t,k)\,dk
\end{equation*}
where the time-correlation function is
\begin{equation*}
\rho(t,k)=e^{-a |k|^{2\beta}t}.
\end{equation*}
Define
$$
V^\varepsilon
(t,x)=\frac{1}{\varepsilon}\,V\left(\frac{t}{\varepsilon^2},x\right).
$$
Then the
time correlation function for $V^\varepsilon (t,x)$ is
\begin{equation*}
\rho(t,k)=\frac{1}{\varepsilon^2}e^{-a
|k|^{2\beta}t/\varepsilon^2}.
\end{equation*}
}

From the general theory of parabolic partial differential equations
\cite{Fr}, for any fixed $\kappa>0, \ep>0$, 
the solution $T^\ep_t(x)$ is a $C^{2+\eta}$-function, with any 
$0<\eta<\alpha-1$.
But the solutions $T^\ep_t$ may lose all the regularity
as  $\kappa\to 0, \ep\to 0$.
So we consider the
weak formulation  of the equation:
\beq
\lan T^\varepsilon, \theta\ran - \lan T_0, \theta \ran &=&
\frac{\kappa}{2}\int_0^t \lan T_s^\vas, \Delta \theta\ran ds
-\frac{1}{\vas}\int_0^t \lan T_s^\vas, \nabla \cdot \left(\theta V
\left(\frac{s}{\vas^2}, \cdot\right)\right)\ran ds
\label{weak}
\eeq
for any test function $\theta \in C^\infty_c(\IR^d)$, the space of
 smooth functions with compact supports. We view
$T^\vas_t$ as distribution-valued processes.
The solutions $\tvas$ can be represented as
\be
\label{1.1}
T^\ep_t(x)=\IM[T_0(\Phi^{t,\ep}_0(x))]
\ee
where $\Phi^{t,\ep}_s$  is  the unique stochastic flow of the SDE
\beq
\label{back-flow}
d\Phi^{t,\ep}_s(x)&=&-\frac{1}{\ep}V(\Phi^{t,\ep}_s(x),\frac{s}{\ep^2})ds+
\kappa^{1/2}dw(t),\quad 0\leq s\leq t\\
\Phi^{t,\ep}_t(x)&=&x.
\eeq

In view of the averaging in the representation (\ref{1.1}) we have
\begin{prop}\label{prop:1}
\[
\|T^\ep_t\|_\infty \leq \|T_0\|_\infty \quad\hbox{a.s.}
\]
\end{prop}
One also has that
\be
\label{23}
\mathbb{E}\{\| T_t^\vas||_p^p\} \leq \|T_0\|_p^p, \quad\forall p\geq 1.
\ee
Indeed, by the spatial homogeneity of the field $V$, the distribution
of $\Phi^{t,\ep}_s(x)$ is the same as the distribution
of $\Phi^{t,\ep}_s(0)+x$ for each fixed $x$. Hence
we have
\[
\IE[\|T^\ep_t\|_p^p]\leq \int\IM\IE[T^p_0(\Phi^{t,\ep}_0(x))]dx
=\IM\IE[\int T^p_0(\Phi^{t,\ep}_0(0)+x)dx]=\|T_0\|^p_p.
\]
Proposition~\ref{prop:1} (resp. (\ref{23})) says 
that, for $T_0\in L^\infty$(resp. $L^p$),
$T^\ep_t$ is almost surely a $L^\infty$(resp. $L^p$)-function  for every $t\geq 0$.

For tightness as well as identification of the limit,
 the following infinitesimal operator  $\cA^\ep$ will play an important role.
Let $V^\vas_t\equiv V(t/\ep^2,\cdot)$. Let $\mathcal{F}_t^\vas$ be the
$\sigma$-algebras generated by $\{V_s^\vas, \, s\leq t\}$  and
$\mathbb{E}_t^\vas$ the corresponding conditional expectation w.r.t. $\cF^\ep_t$.
Let $\cM^\ep$ be the space of measurable function adapted to $\{\cF^\ep_t, \forall t\}$ 
such that $\sup_{t<t_0}\IE|f(t)|<\infty$.
We say $f(\cdot)\in \cD(\cA^\ep)$, the domain of $\cA^\ep$, and $\cA^\ep f=g$
if $f,g\in \cM^\ep$ and for
$f^\delta(t)\equiv\delta^{-1}[\IE^\ep_t f(t+\delta)-f(t)]$
we have
\beqn
\sup_{t,\delta}\IE|f^\delta(t)|&<&\infty\\
\lim_{\delta\to 0}\IE|f^\delta(t)-g(t)|&=&0,\quad\forall t.
\eeqn
 For $f(t)=\phi(\lan T_t^\vas, \theta\ran),
 f'(t)=\phi'(\lan T_t^\vas, \theta\ran),
 \forall \phi\in C^3_c(\IR)$ (i.e. $C^3$-function with a compact support)
we have the following expression 
from (\ref{weak}) and the chain rule 
\beq
\label{gen}
 \cA^\vas
f(t)
&=& \frac{\kappa}{2} f'(t) \lan T_t^\vas,
\Delta \theta\ran - \frac{1}{\vas} f'(t) \lan T_t^\vas,
\cv(\theta)\ran
\eeq
where 
\be
\label{nu}
\cv(\theta) \equiv\nabla\cdot [\theta V^\ep_t].
\ee
 A main property of $\cA^\ep$ is
that 
\be
\label{12.2}
f(t)-\int^t_0 \cA^\ep f(s) ds\quad\hbox{is a  $\cF^\ep_t$-martingale},
\quad  f\in \cD(\cA^\ep).
\ee
Also,
\be
\label{mart}
\IE^\ep_t f(s)-f(t)=
\int^s_t \IE^\ep_t \cA^\ep f(\tau) d\tau\quad \forall s>t \quad\hbox{a.s.}
\ee
(see \cite{Kur}).
We can view $\tvas$ as the distribution-valued
stochastic solutions to the martingale problem
(\ref{12.2}).

Likewise we formulate the solutions for the Kraichnan's model
(\ref{kr}) as the solutions to the corresponding martingale problem.
We will first describe the limiting martingale problem for Theorem~1
and then discuss the changes due to $\ell_1\to 0$ in Theorem~2.
We rewrite (\ref{kr}) as an It\^{o}'s SDE
\be
dT_t=
\left(\frac{\kappa_0}{2}\Delta +
\frac{1}{a}\cB\right)
T_t\,dt+\sqrt{2}a^{-1/2}\n T_t\cdot dW^{(1)}_t 
\label{14.2}
\ee
where $W_t^{(1)}(x)$ is the Brownian vector field with the spatial
covariance 
\[
\Gamma^{(1)}(x-y)=\int \exp{(ik\cdot(x-y))}\cE(\alpha+\beta,k)|k|^{1-d}dk
\]
and the operator $\cB=\cB_1+\cB_2$ is given by 
\beq
\label{43'}
\cB_1\phi&=&
\sum_i
\left[\frac{\partial}{\partial x_i}\Gamma_{ij}^{(1)}(0)\right]
\frac{\partial\phi}{\partial x_j}
\\
\cB_2\phi&=&\sum_{i,j}\Gamma_{ij}^{(1)}(0)
\frac{\partial^2 \phi}{\partial x_i\partial x_j}.
\label{44'}
\eeq
Eq. (\ref{14.2}) can be
\commentout{
 understood as
\be
\label{13.2'}
 \lan T_t,\theta\ran=
 \int_0^t \left[\frac{\kappa}{2}\lan T_s,\Delta\theta\ran +
 \frac{1}{a}\lan T_s,\cB^*\theta\ran\right]\,ds 
 +\sqrt{\frac{1}{a}}\int_0^t \n T_s\cdot dW^{(1)}_s
 \ee
 or that
}
formulated as the martingale problem: Find a measure $\IP$ 
(of $T_t$) on the subspace of $D([0,t_0);L^\infty_{w^*}(\IR^d))$
whose elements have a given initial data in $L^\infty_{w^*}(\IR^d)$
such that
 \beq
 \label{13.2}
 &&f( \lan T_t,\theta\ran)-\int_0^t
 \bigg\{f'( \lan T_t,\theta\ran)\left[\frac{\kappa_0}{2}\lan
 T_s,\Delta\theta\ran+\frac{1}{a}\lan T_s,\cB^*\theta\ran\right]
  + \frac{1}{a}f''(\lan T_t,\theta\ran) 
  \lan\theta, \cK^{(1)}_{T_s}\theta\ran\bigg\}\,ds\\
\nonumber&&\hbox{{\em is a martingale w.r.t. the filtration of a cylindrical
Wiener process, for each} $f\in C^3_c(\IR)$}
\eeq
where $\cB^*$ is the adjoint of $\cB$ and
$\cK^{(1)}_{T_t}$ is a positive-definite operator given formally as
\be
\label{15.2}
\cK^{(1)}_{T_t}\theta=
\int\theta(y) \nabla T_t(x)\cdot \Gamma^{(1)}(x-y)\n T_t(y)
\,dy
\ee
such that 
\beq
\lan\theta_1,\cK^{(1)}_{\phi}\theta_2\ran& =&
\iint\phi(x)\phi(y)G^{(1)}_{\theta_1,\theta_2}(x,y)\,dx\,dy\\
G^{(1)}_{\theta_{1},\theta_{2}}&\equiv&
\sum_{i,j}\frac{\partial^2}{\partial
x_i
\partial
y_j}\left[\theta_1(x)\theta_2(y)\Gamma_{ij}^{(1)}(x-y)\right].
\eeq

When $\ell_1\to 0$ (Theorem~2)
$\Gamma^{(1)}$ in the preceding discussion should
be replaced by
\be
\label{55}
\bar{\Gamma}^{(1)}(x-y)=\lim_{\ell_1\to 0}\Gamma^{(1)}(x-y)
\ee
and all objects (such as $\cB, G_{\theta_1,\theta_2}, \cK_{T_t}^{(1)}$)
related to $\Gamma^{(1)}$  
should be replaced accordingly (by $\bar{\cB},\bar{G}_{\theta_1,\theta_2},
\bar{\cK}_{T_t}^{(1)}$).
In particular,  
$\bar{\cB}_1$ is well-defined only for $\alpha+\beta>3/2$ in general
for compressible flows. Namely, the martingale problem
(\ref{13.2}) is not well defined in the compressible case
unless the limiting Brownian velocity has a spatial Hurst exponent
which is bigger than $1/2$.

In the case of divergence-free vector fields,
$\bar{\cB}_1=0$ and
\beq
\label{56}
\bar{\cB}\phi&=&\sum_{i,j}\bar{\Gamma}_{ij}^{(1)}(0)
\frac{\partial^2 \phi}{\partial x_i\partial x_j}.
\eeq
Also, 
\[
\bar{G}^{(1)}_{\theta_{1},\theta_{2}}\equiv
\sum_{i,j} \bar{\Gamma}_{ij}^{(1)}(x-y)
\frac{\partial \theta_1(x)}{\partial x_i}
\frac{\partial \theta_2(y)}{\partial y_j}.
\]

\subsection{Uniqueness of the limiting Kraichnan model}
When $\lim_{\ep\to0}\ell_1>0$ the limit Brownian velocity field
is spatially smooth
and generates a unique flow of diffeomorphisms on $\IR^d$ 
(\cite{Bax}, \cite{BH}) from which it follows the uniqueness of
the martingale solution.

When $\lim_{\ep\to0}\ell_1=0$ the limiting velocity field
is only spatially H\"{o}lder continuous and we establish the uniqueness
of the martingale solution by proving the uniqueness
of the $n$-point correlation function
\[
F^t_n(x_1,x_2,x_3,...,x_n)\equiv\IE_{T_0}\lt[T_t(x_1)T_t(x_2)\cdots T_t(x_n)\rt].
\]
The evolution of the $n$-point correlation function is given by 
a weakly continuous (hence strongly continuous)
sub-Markovian semigroup 
on $L^p(\IR^{nd})$,
$\forall p\in (1,\infty)$ whose generator.
can be deduced by taking the test function
$f(r)=r^n$ in the martingale formulation:
\be
\label{ndiff}
 \cL_n \Phi(x_1,\cdots,x_n)\equiv \frac{\kappa_0}{2}\sum_{j=1}^n
 \Delta_{x_j} \Phi+\frac{1}{a}\sum_{i,j=1}^n \bar{\Gamma}^{(1)}(x_i-x_j):
 \nabla_{x_i}\nabla_{x_j}\Phi,\quad\Phi\in C^\infty_c(\IR^{nd}),\,\,
 \kappa_0\geq 0.
 \ee
 Note that the symmetric operator $\cL_n$ (\ref{ndiff})
 is an essentially self-adjoint positive operator  on $C_c^\infty(\IR^{nd})$,
 which then induces a {\em unique} symmetric Markov semigroup
 of contractions on $L^2(\IR^{nd})$.
 The essential self-adjointness
 is due to the sub-Lipschitz growth of the square-root
 of $\bar{\Gamma}^{(1)}(x_1- x_2)$ at large $|x_1|, |x_2|$
 (hence no escape to infinity) \cite{Da}.

In the sequel we will adopt the following notation 
\[
f(t)\equiv f(\lan T_t^\vas, \theta\ran),\quad
 f'(t)\equiv f'(\lan T_t^\vas, \theta\ran),\quad
f''(t)\equiv f''(\lan T_t^\vas, \theta\ran),\quad
f'''(t)\equiv f'''(\lan T_t^\vas, \theta\ran) 
\quad\forall f\in C_c^3(\IR).
\]
Namely, the prime stands for the differentiation w.r.t. the original argument
($\lan T_t^\vas, \theta\ran$ not $t$)
of $f, f'$ etc.

\section{Proof of Theorem~1}
The proofs are a refinement of that of \cite{CF}
to deal with the wave-number dependence
of the correlation time and the lack of spatial regularity
in the velocity fields.
For the reader's convenience, we will repeat some of
the calculations in \cite{CF} and refer the reader to
\cite{Ku} for the full exposition of the perturbed test
function method used here. The perturbed test function
method is initiated in \cite{PSV}.

\subsection{Tightness}
A family of distribution-valued right-continuous with left 
limits processes $\{T^\ep, 0<\ep<1\} $ is 
tight if and only if the family of real-valued,
right-continuous with left limits processes
$\{\lan T^\ep, \theta\ran, 0<\ep <1\}$ is tight for all $\theta\in C^\infty_c(\IR^d)$. 
We use the tightness  criterion of \cite{Ku}
(Chap. 3, Theorem 4), namely, 
we will prove:
Firstly,
\be
\label{trunc}
\lim_{N\to \infty}\limsup_{\ep\to 0}\IP\{\sup_{t<t_0}|\lan T^\ep, \theta\ran|
\geq N\}=0,\quad\forall t_0<\infty.
\ee
Secondly, for  each $f\in C^3_c(\IR)$  
there is a sequence
$f^\ep(t)\in\cD(\cA^\ep)$ such that for each $t_0<\infty$
$\{\cA^\ep f^\ep (t), 0<\ep<1,0<t<t_0\}$ 
is uniformly integrable and
\be
\label{19}
\lim_{\ep\to 0} \IP\{\sup_{t<t_0} |f^\ep(t)-
f(\lan T^\ep, \theta\ran) |\geq \delta\}=0,\quad \forall \delta>0.
\ee
Then it follows that the laws of
$\{\lan T^\ep, \theta\ran, 0<\ep <1\}$ are tight in the space
$L^\infty_{w^*}(\IR^d)$.

Condition (\ref{trunc}) is satisfied as a result of Proposition~1. 
Let
\[
f_1^\vas (t)\equiv \frac{1}{\vas}\int_t^\infty
\mathbb{E}_t^\vas\, f'(t) \lan T_t^\vas,\cV^\ep_s(\theta)\ran\,ds
\]
be the 1-st perturbation of $f(t)$. Using the spectral representation
\begin{equation}
\begin{split}
\label{cond-exp}
\mathbb{E}_t^\vas\, V_s^\vas = \int e^{ix\cdot k}
e^{-a |k|^{2\beta}|s-t|\vas^{-2}} \widehat{V}_t^\vas (dk),
 \quad \forall \, s\geq t,
\end{split}
\end{equation}
we obtain
\begin{equation}
\label{1st}
f_1^\vas (t)= 
\frac{\vas}{a}
f'(t) \lan T_t^\vas,\cvtil(\theta)\ran
\end{equation}
with
\bea
\label{cvtil}
\cvtil(\theta) &=&\nabla\cdot [\theta \tilde{V}^\ep_t]\\
\tilde{V}_t^\vas 
&\equiv&\tilde{V} \left(\frac{t}{\vas^2}, \cdot\right)\equiv
\ep^{-2}\int_t^\infty
\IE_t^\vas\, V_s^\vas \, ds
\eea
where $\tilde{V}$ has the power spectrum 
$\cE(\alpha+2\beta,k)$.

\begin{prop}\label{prop:2}
\begin{enumerate}
$$\lim_{\ep\to 0}\sup_{t<t_0} \mathbb{E} |f_1^\vas(t)|=0,\quad
\lim_{\ep\to 0}\sup_{t<t_0} |f_1^\vas(t)|= 0
\quad \hbox{in probability}$$.
\end{enumerate}
\end{prop}

\begin{proof}
By Proposition \ref{prop:1} we have
\be
\label{1.2}
\mathbb{E}[|f_1^\vas(t)|]\leq \frac{\vas}{a} \|f'\|_\infty
\|T_0\|_\infty 
\left[
\|\theta\|_\infty\int_{|x|\leq M} \IE|\tilde{V}^\ep_t(x)|\,\,dx+
\|\n\theta\|_\infty\int_{|x|\leq M}\IE|\n\cdot\tilde{V}^\ep_t|dx\right]
\ee
and
\beq
\label{1.3}
\lefteqn{\sup_{t< t_0} |f_1^\vas(t)|}\\
 &\leq& \frac{\vas}{a}
\|f'\|_\infty  \|T_0\|_\infty\left[
\|\theta\|_\infty\sup_{t<t_0}\int_{|x|\leq M} |\tilde{V}^\ep_t(x)|\,\,dx+
\|\n\theta\|_\infty\sup_{t<t_0}\int_{|x|\leq M}|\n\cdot\tilde{V}^\ep_t|dx\right].
\nonumber
\eeq
By the temporal stationarity of $\tilde{V}^\ep_t$ we can replace
$\IE|\tilde{V}^\ep_t(x)|, \IE|\nabla\cdot\tilde{V}^\ep_t(x)|$ in (\ref{1.2}) by 
$\IE|\tilde{V}(0,x)|, \IE|\nabla\cdot\tilde{V}(0,x)|$.
By the Gaussianity, temporal stationarity and spatial
homogeneity of $\tilde{V}$, we 
can replace $\sup_{t<t_0}\int_{|x|\leq M} |\tilde{V}^\ep_t(x)|\,\,dx$
in (\ref{1.3}) by
\begin{equation}
\label{1.4}
M^d\sup_{\substack{|x|\leq M\\t\leq t_0}} \left|\tilde{V}\left(\frac{t}{\vas},x\right)\right|\leq
C\log \left[\frac{M^dt_0}{\vas^2} \right]= o\lt(\frac{1}{\ep}\rt)
\end{equation}
with a random constant $C$ possessing a distribution with
a finite moment (Indeed, a Gaussian-like tail by Chernoff's bound).
A similar inequality holds for $\nabla\cdot\tilde{V}$.
Proposition~\ref{prop:2}
now follows from (\ref{1.2}), (\ref{1.3}) and (\ref{1.4}).
\end{proof}

Set $f^\ep(t)=f(t)-f^\ep_1(t)$.
A straightforward calculation yields
\begin{align*}
\begin{split}
 \cA^\vas f_1^\vas &=-\frac{k\vas}{2a}f''(t)\lan
T_t^\vas,\Delta\theta \ran\lan T_t^\vas,
\cvtil(\theta) \ran +\frac{k\vas}{2a}f'(t)\lan
 T_t^\vas,\Delta\cvtil(\theta)\ran\\
&\quad +
\frac{1}{a}f''(t)\lan T_t^\vas,\cv(\theta)\ran\lan
T_t^\vas,\cvtil(\theta)\ran  -
\frac{1}{a} f'(t)\lan\tvas,\cv(\cvtil(\theta))\ran \\
&\quad +
\frac{1}{\vas}f'(t)\lan\tvas,\cv(\theta)\ran
\end{split}
\end{align*}
and, hence
\begin{align}
\label{25'}
\begin{split}
\cA^\vas f^\ep(t)
&=\frac{\kappa}{2}f'(t)\lan \tvas,\Delta\theta \ran - \frac{1}{a}f'(t)\lan\tvas,
\cv(\cvtil(\theta))\ran
-\frac{1}{a}f''(t)\lan\tvas, \cv(\theta)\ran\lan\tvas,
\cvtil(\theta)\ran \\
&\quad +
\frac{\kappa\vas}{2a}\left[f''(t)\lan\tvas,\Delta\theta\ran\lan\tvas,
\cv(\theta)\ran
 -f'(t)\lan\tvas,
\Delta\cvtil(\theta)\ran\right] \\
&=A_1^\vas(t)+A_2^\vas(t)+A_3^\vas(t)+A_4^\vas(t)
\end{split}
\end{align}
where $A_2^\vas(t)$ and $A_3^\vas(t)$ are the $O(1)$ statistical coupling
terms.

For the tightness criterion stated in the beginnings of the section,
it remains to show
\begin{prop}
$\{\cA^\ep f^\ep\}$ are uniformly integrable and
$$\lim_{\ep\to 0}\sup_{t<t_0}\IE|A^\ep_4(t)|=0$$.
\end{prop}

\begin{proof}
We show that $\{A^\ep_i\}, i=1,2,3,4$ are uniformly integrable. 
To see this, we have the following estimates.
\bean
|A_1^\vas(t)|=\frac{\kappa}{2}\lt|f'(t)\lan \tvas,\Delta\theta \ran\rt|
\leq\frac{\kappa}{2}\|f'\|_\infty\|T_0\|_\infty\|\Delta \theta\|_1
\eean
Thus $A^\ep_1$ is uniformly integrable since it is uniformly bounded.
\beqn
|A_2^\vas(t)|&=&\frac{1}{a}\lt|f'(t)\lan\tvas, 
\cv(\cvtil(\theta))\ran\rt|\\
&\leq& \frac{C}{a}\|f'\|_\infty\|T_0\|_\infty
\left[
\int_{|x|<M}|{V}^\ep_t|^2\,dx+
\int_{|x|<M}|\n\cdot {V}^\ep_t|^2\,dx\right]^{1/2}\times\\
&&\quad\left[\int_{|x|<M}|\tilde{V}^\ep_t|^2\,dx+
\int_{|x|<M}|\n\cdot\tilde{V}^\ep_t|^2\,dx +
\int_{|x|<M}|\n\n\cdot\tilde{V}^\ep_t|^2\,dx\right]^{1/2}.
\eeqn
Thus $A^\ep_2$ is  uniformly integrable in view of
the uniform boundedness of the 4-th moment of
$V^\ep_t, \tilde{V}^\ep_t$ and their spatial derivatives
due to Gaussianity and the ultraviolet cutoff $\ell_1>0$.
\beqn
|A_3^\vas(t)|&=& \frac{1}{a}\lt|f''(t)\lan\tvas,\cv(\theta)\ran\lan\tvas,
\cvtil(\theta)\ran\rt|\\
&\leq& \frac{C}{a}\|f'\|_\infty\|T_0\|_\infty^2
\left[\int_{|x|<M}|{V}^\ep_t|^2\,dx+
\int_{|x|<M}|\n\cdot {V}^\ep_t|^2\,dx+
\int_{|x|<M}|\tilde{V}^\ep_t|^2\,dx+
\int_{|x|<M}|\n\cdot\tilde{V}^\ep_t|^2\,dx\right].
\eeqn
Thus $A^\ep_3$ is  uniformly integrable for the similar reason that
 $A^\ep_2$ is  uniformly integrable.
 \beq
 |A^\ep_4|
  \nonumber
 &=&\frac{\kappa\vas}{2a}
 |f''(t)\lan\tvas,\Delta\theta\ran\lan\tvas,
 \cvtil(\theta)\ran
  - f'(t)\lan\tvas,\Delta\cvtil(\theta)\ran\\
  \nonumber
  &\leq& \frac{C\kappa\vas}{2a}
  \left[\|f''\|_\infty\|T_0\|_\infty^2
\left[\int_{|x|<M}|\tilde{V}^\ep_t|^2\,dx+
\int_{|x|<M}|\n\cdot\tilde{V}^\ep_t|^2\,dx\right]^{1/2}\right.
 +
\|f'\|_\infty\|T_0\|_\infty \times\\ &&\quad\left.\left[
\int_{|x|<M}|\tilde{V}^\ep_t|^2\,dx+
\int_{|x|<M}|\n\tilde{V}^\ep_t|^2\,dx
+\int_{|x|<M}|\n^2\tilde{V}^\ep_t|^2\,dx+
\int_{|x|<M}|\n^2\n\cdot\tilde{V}^\ep_t|^2\,dx\right]^{1/2}
\right].
 \label{1.10}
  \eeq
  Due to the fixed cutoff $\ell_1>0$, the higher derivatives
  of $\tilde{V}^\ep_t$ do not cause any difficulty and
  they all have uniformly bounded, say, the 4-th moments. Hence
  $A^\ep_4$ is uniformly integrable.
Clearly $$\lim_{\ep\to 0}\sup_{t<t_0}\IE|A^\ep_4(t)|=0.$$
  \end{proof}

\subsection{Identification of the limit}
Once the tightness is established we can use another result 
 in \cite{Ku} (Chapter 3, Theorem 2) to identify the limit.
Let $\cA$ be a diffusion or jump diffusion operator such that
there is a unique solution $\omega_t$ in the subspace of $
D([0,t_0); L^\infty_{w^*}(\IR^d)), \forall t_0<\infty,$
 whose elements have the given initial data in
 $L^\infty_{w^*}(\IR^d)$ such that 
\be
\label{38}
f(\omega_t)-\int^t_0\cA f(\omega_s)\,ds
\ee
is a martingale. 
We shall show that for each $f\in C^3_c(\IR)$ there exists $f^\ep\in \cD(\cA^\ep)$
such that
\beq
\label{38.2}
\sup_{t<t_0,\ep}\IE|f^\ep(t)-f(\lan T^\ep_t,\theta\ran)|&<&\infty\\
\label{39.2}
\lim_{\ep\to 0}\IE|f^\ep(t)-f(\lan T^\ep_t,\theta\ran)|&=&0,\quad \forall t<t_0\\
\label{40.2}
\sup_{t<t_0,\ep}\IE|\cA^\ep f^\ep(t)-\cA f(\lan T^\ep_t,\theta\ran)|&<&\infty\\
\lim_{\ep\to 0}\IE|\cA^\ep f^\ep(t)-\cA f(\lan T^\ep_t,\theta\ran)|&=&0,\quad
\forall t<t_0.
\label{42.2}
\eeq
Then the aforementioned theorem implies that any tight processes
$\lan T^\ep_t,\theta\ran$ converges
in law to the unique process generated by $\cA $.
As before we adopt the notation $f(t)=f(\lan T^\ep_t,\theta\ran)$.

For this purpose,
we introduce the next perturbations $f_2^\ep, f_3^\ep$.
Let
\bea
\label{40.3}
A_2^{(1)}(\phi) &\equiv&\lan\theta,\cK^{(1)}_\phi\theta\ran\\
A_3^{(1)}(\phi)&\equiv&\lan\phi, \IE\lt[\cv(\cvtil(\theta))\rt]\ran
\eea
where the positive-definite operator $\cK^{(1)}_{\phi} $ is defined
in (\ref{15.2}).
It is easy to see that
\begin{align}
A_2^{(1)}(\phi)&=\mathbb{E}\lt[\lan\phi, \cv(\theta)\ran
\lan\phi, \cvtil(\theta)\ran\rt]\\
A_3^{(1)}(\phi)&= \lan\cB\phi,\theta\ran
\label{41.2}
\end{align}
where the operator $\cB=\cB_1+\cB_2$ is given by
(\ref{43'}) and (\ref{44'}).

Define
\begin{align*}
f_2^\vas(t) &\equiv
\frac{1}{a}f''(t)\int_t^\infty \mathbb{E}_t^\vas
\lt[\lan\tvas, \cV^\ep_s(\theta)\ran\lan\tvas,\tilde{\cV}^\ep_s(\theta)\ran -A^{(1)}_2(\tvas)\rt]\,ds
\\
f_3^\vas(t) &\equiv \frac{1}{a}f'(t)\int_t^\infty \mathbb{E}_t^\vas
\lt[\lan\tvas, \cV^\ep_s(\tilde{\cV}^\ep_s(\theta))\ran-A^{(1)}_3(\tvas)\rt]\,ds.
\end{align*}
Let
\begin{align*}
G^{(2)}_{\theta_{1},\theta_{2}}&\equiv\sum_{i,j}\frac{\partial^2}{\partial
x_i
\partial
y_j}\left[\theta_1(x)\theta_2(y)\Gamma_{ij}^{(2)}(x-y)\right]\\
\lan\theta_1,\cK^{(2)}_\phi \theta_2\ran &\equiv
\iint\phi(x)\phi(y)G^{(2)}_{\theta_1,\theta_2}(x,y)\,dx\,dy
\end{align*}
where the covariance function $\Gamma^{(2)}(x-y)\equiv
\IE\lt[\veptil(x)\otimes\veptil(y)\rt]$ has the spectral density
$\cE(\alpha+2\beta,k)$, and let
\begin{equation*}
\begin{split}
A_2^{(2)}(\phi)&\equiv \lan \theta, \cK^{(2)}_\phi\theta\ran\\
A_3^{(2)}(\phi)&\equiv \lan \phi,\IE\lt[\cvtil(\cvtil(\theta))\rt]\ran.
\end{split}
\end{equation*}

Noting that
\begin{equation}
\label{cond1}
\begin{split}
&\mathbb{E}_t^\ep[V_s^\vas(x)\otimes\tilde{V}_s^\vas(y)] \\
&=\int e^{i(x-y)\cdot
k}\hat{V}_t^\vas(dk)\otimes
\hat{\tilde{V}}_t^\vas(dk)e^{-2a{|k|}^{2\beta}|s-t|\vas^{-2}}\\
&\quad +
\int e^{i(x-y)\cdot
k}
\left[1-e^{-2a{|k|}^{2\beta}|s-t|\vas^{-2}}\right] \cE(\alpha+\beta,k)
\,dk
\end{split}
\end{equation}
\commentout{
and
\begin{equation}
\label{cond2}
\begin{split}
&\mathbb{E}_t^\ep[\tilde{V}_s^\vas(x)\otimes\tilde{V}_s^\vas(y)] \\
&=\int e^{i(x-y)\cdot
k}\hat{\tilde{V}_t^\vas}(dk)\otimes
\hat{\tilde{V}_t^\vas}(dk)e^{-2a{|k|}^{2\beta}|s-t|\vas^{-2}}\\
&\quad +
\int e^{i(x-y)\cdot
k}
\left[1-e^{-2a{|k|}^{2\beta}|s-t|\vas^{-2}}\right] \cE(\alpha+2\beta,k)
\,dk.
\end{split}
\end{equation}
}
we then have
\be
\label{1.44}
f_2^\vas(t)=\frac{\vas^2}{2a^2}f''(t)\left[{\lan
\tvas,\cvtil(\theta)\ran}^2-A_2^{(2)}(\tvas)\right]
\ee
and similarly
\be
\label{1.45}
f_3^\vas(t)=\frac{\vas^2}{2a^2}f'(t)
\lt[\lan\tvas,\cvtil(\cvtil(\theta))\ran-A_3^{(2)}(\tvas)\rt].
\ee

In view of the prefactor $\ep$ in (\ref{1.44}) and
(\ref{1.45}) and  the fact that all terms involved
are regular and uniformly bounded, we have 
\begin{prop}\label{prop:5}
$$ \lim_{\ep\to 0}\sup_{t<t_0} \mathbb{E}|f_2^\vas(t)|=0,\quad \lim_{\ep\to 0}\sup_{t<t_0}
\mathbb{E}|f_3^\vas(t)|=0. $$
\end{prop}
The proof of Proposition~\ref{prop:5} is analogous to
that of Proposition~2.

We have
\beqn
\cA^\vas f_2^\vas(t)&=&\frac{1}{a}
f''(t)\left[
\lan\tvas, \cv(\theta)\ran\lan\tvas,\cvtil(\theta)\ran - A^{(1)}_2(\tvas)\right]
+ R_2^\vas(t)\\
\cA^\vas f_3^\vas(t)&=&\frac{1}{a}
f'(t)\left[\lan
\tvas,\cv(\cvtil(\theta))\ran - A^{(1)}_3(\tvas)\right]+
R_3^\vas(t)
\eeqn
with
\beq
R_2^\vas(t)& =& f'''(t)\left[
\frac{\ep^2\kappa}{4a^2}\lan \tvas, \Delta\theta\ran-\frac{\ep}{2a^2}
\lan \tvas, \cv(\theta)\ran\right]
\left[\lan \tvas,\cvtil(\theta)\ran^2-A^{(2)}_2(\tvas)\right]\nonumber\\
&&\quad +
f''(t) \lan\tvas,\cvtil(\theta)\ran\left[
\frac{\kappa\ep^2}{2a^2}\lan \tvas,\Delta\cvtil(\theta)\ran-
\frac{\ep}{a^2}\lan \tvas,\cv(\cvtil(\theta))\ran\right]\nonumber\\
&&\quad -
f''(t)\left[\frac{\kappa\ep^2}{4a^2} \lan \tvas,\Delta G^{(2)}_{\theta}\tvas\ran
-\frac{\ep}{a^2}\lan \tvas,\cv(G^{(2)}_{\theta}\tvas)\ran\right]
\label{36}
\eeq
where in (\ref{36})
$G_\theta^{(2)}$ denotes the operator
\[
G_\theta^{(2)}\phi\equiv \int G^{(2)}_{\theta,\theta}(x,y)\phi(y)\,dy,
\]
and similarly
\beqn
R^\ep_3(t)&=&f''(t)\left[\frac{\kappa\ep^2}{4a^2}\lan\tvas,\Delta\theta\ran
-\frac{\ep}{2a^2}\lan\tvas,\cv(\theta)\ran\right]
\left[\lan\tvas,\cvtil(\cvtil(\theta))\ran-A^{(2)}_3(\tvas)\right]\\
&&\quad+f'(t)\left[\frac{\kappa\ep^2}{4a^2}
\lan\tvas,\Delta\cvtil(\cvtil(\theta))\ran-
\frac{\ep}{2a^2}\lan\tvas,\cv(\cvtil(\cvtil(\theta)))\ran\right]\\
&&\quad-f'(t)\left[\frac{\kappa\ep^2}{4a^2}\lan\tvas,\Delta
\IE[\cvtil(\cvtil(\theta))]\ran+\frac{\ep}{2a^2}\lan\tvas,\cv(\IE[\cvtil(\cvtil(\theta))])\ran\right].
\eeqn

Now all terms appearing in $ R^\ep_2(t)$ and $R^\ep_3(t)$ are
regular and uniformly bounded, we easily have
\begin{prop}\label{prop:6}
\[
\lim_{\ep\to 0}\sup_{t<t_0} \mathbb{E} |R_2^\vas(t)|=0,\quad \lim_{\ep \to 0}
\sup_{t<t_0} \mathbb{E}
|R_3^\vas(t)|=0.
\]
\end{prop}

Set
\[
R^\vas(t) = A_4^\vas(t) + R_2^\vas(t) + R_3^\vas(t).
\]
It follows from Propositions 3 and 5 that
\[
\lim_{\ep \to 0}\sup_{t<t_0}\IE|R^\ep(t)|=0.
\]
Recall that
\beqn
M_t^\vas(\theta)&=&f^\ep(t)-\int^t_0 \cA^\ep f^\ep(s)\,ds\\
&=& f(t)-f_1^\vas(t)+f_2^\vas(t)+f_3^\vas(t)
-
\int_0^t\frac{\kappa}{2}f'(t)\lan\tvas,\Delta\theta\ran\,ds\\
&& - \int_0^t\frac{1}{a}\left[f''(t)(\lan T_s^\vas,\theta\ran)
A_2^{(1)}(T_s^\vas)+f'(t) A_3^{(1)}(T_s^\vas)\right]\,ds -\int_0^t
R^\vas(s)\,ds
\eeqn
is a martingale. Now that (\ref{38.2})-(\ref{42.2}) are satisfied
we can identify the limiting martingale to be
\begin{equation}
M_t(\theta)=f(t)-\int_0^t
\bigg\{f'(s)\left[\frac{\kappa_0}{2}\lan
T_s,\Delta\theta\ran+\frac{1}{a}{A}_3^{(1)}(T_s)\right]
 + \frac{1}{a}f''(s){A}_2^{(1)}(T_s)\bigg\}\,ds.
 \label{37}
\end{equation}

Since $\lan\tvas,\theta\ran$ is uniformly bounded
\[
\lt|\lan\tvas,\theta\ran\rt|\leq \|T_0\|_\infty{\|\theta\|}_1
\]
we have the convergence of the second moment
\[
\lim_{\ep\to 0}
\IE\left\{
{\lan\tvas,\theta\ran}^2\right\}=\mathbb{E}\left\{ {\lan
T_t,\theta\ran}^2\right\}.
\]
Use $f(r) =r$ and $r^2$ in (\ref{37})
$$ M_t^{(1)}(\theta)=\lan T_t,\theta\ran -
\int_0^t \left[\frac{\kappa_0}{2}\lan T_s,\Delta\theta\ran +
\frac{1}{a}A_3^{(1)}(T_s)\right]\,ds $$ is a martingale with the
quadratic variation
$$
\left[M^{(1)}(\theta),M^{(1)}(\theta)\right]_t=
\frac{2}{a}\int_0^tA_2^{(1)}(T_s)\,ds=
\frac{2}{a}\int_0^t
\lan\theta,\cK^{(1)}_{T_s}\theta\ran\,ds. $$ 
Therefore, $$
M_t^{(1)}=\sqrt{\frac{2}{a}}\int_0^t \sqrt{\cK^{(1)}_{T_s}}dW_s $$ where
$W_s$ is a cylindrical Wiener process (i.e.
$dW_t(x)$ is a space-time white noise field)
and $\sqrt{\cK^{(1)}_{T_s}}$ is the square-root of
the positive-definite operator given  in (\ref{15.2}).
From (\ref{40.3}) and (\ref{41.2}) we see that the limiting process $T_t$
is the (assumed unique) distributional solution
to the martingale problem (\ref{13.2}) of
the It\^{o}'s equation 
\beqn
dT_t&=&
\left(\frac{\kappa_0}{2}\Delta +
\frac{1}{a}\cB\right)
T_t\,dt+\sqrt{2a^{-1}\cK^{(1)}_{T_t}}\,dW_t\\
&=&\left(\frac{\kappa_0}{2}\Delta +
\frac{1}{a}\cB\right)
T_t\,dt+\sqrt{2}a^{-1/2}\n T_t\cdot dW^{(1)}_t 
\eeqn
where the operator $\cB=\cB_1+\cB_2$ is given by (\ref{43'})-(\ref{44'})
 and
$W_t^{(1)}$ is the Brownian vector field with the spatial
covariance $\Gamma^{(1)}(x-y)$.

\section{Proof of Theorem~2}

\commentout{
When the velocity field is incompressible and has
sufficiently high order moments, the standard theory
of parabolic PDE (see \cite{LU}, Chap. III, Theorem 7.2)
has the energy identity
\be
\label{45}
\int |\tvas(x)|^2 \,dx+\kappa\int^t_0\int |\n \tvas|^2(x)\,dx\,ds
=\int|T_0(x)|^2.
\ee
Note that  the equality (\ref{45}) holds even for $\kappa=0$.
Also, the stochastic flow $\Phi^{t,\ep}_s(x)$
(\ref{back-flow}) is almost surely well-defined
for $\kappa=0$ since the velocity is smooth with $\ell_1=\ell_1(\ep)>0$
and satisfies the sub-linear growth condition almost surely. In this
case, $\Phi^{t,\ep}_s(x)$ is a Lebesgue-measure preserving diffeomorphism
for each $t, s,\ep>0$. We have the representation
\[
\tvas(x)=T_0(\Phi^{t,\ep}_0(x))
\]
in view of which
Proposition~1 clearly holds. 
The proof of Proposition~2 is the same except that all the terms
containing $\n\cdot\veptil$ vanish.
}

The argument is the same as before except with
\beqn
\cv(\theta)&\equiv& \vep\cdot\n\theta\\
\cvtil(\theta)&\equiv& \veptil\cdot\n\theta,
\eeqn
instead of (\ref{nu}) and (\ref{cvtil}), because of the
incompressibility of the velocity fields.
Also, all the terms containing $\n\cdot\vep$ and $\n\cdot\veptil$
vanish.
\commentout{
It is important to note that while $\vep$ loses differentiability
as $\ell_1\to 0$,  $\veptil$ is almost surely a $C^{1,\eta}$-function
in the limit with
\[
0<\forall \eta<\alpha+2\beta-2
\]
and has uniformly bounded local $W^{1,p}$-norm, $p\geq 1$.
}

The most severe term to occur 
in the argument for tightness (in the expression for $A_4^\ep$)
is 
\[
\frac{\kappa\ep}{2a}\lt|f'(t)\lan
\tvas,\Delta\cvtil(\theta)\ran\rt|
\]
whose second moment can be bounded as
\beq
\nonumber
\lefteqn{\frac{\kappa\ep}{2a}\sqrt{\IE\lt|f'(t)\lan
\tvas,\n^2\cvtil(\theta)\ran\rt|^2}}\\
\nonumber
&\leq&C_1\frac{\kappa\ep}{2a}\|f'\|_\infty
\|T_0\|_\infty\lt(\int_{|x|<M}\IE\lt[|\Delta\veptil|^2\rt]\,dx\rt)^{1/2}\\
&\leq& C_2\kappa\ep\times 
\left\{\begin{array}{ll}
\ell_1^{\alpha+ 2\beta-3},&\hbox{for}\,\,\alpha+2\beta <3\\
\sqrt{\log{(1/\ell_1)}},&\hbox{for}\,\,\alpha+2\beta=3\\
1,&\hbox{for}\,\, \alpha+2\beta>3.
\end{array}\rt.
\label{39}
\eeq
Other possibly divergent terms occurring in identifying the limit
can be controlled similarly. For instance,
the most severe term without the prefactor $\kappa$
occurs in $R^\ep_3(t)$ and  can be
bounded as
\beq
\ep\IE\left|\lan \tvas,\cv(\cvtil(\cvtil(\theta)))\ran\right| &\leq&
\ep\|T_0\|_\infty
\IE^{1/2}|\cv(\cvtil(\cvtil(\theta)))|^2 \nonumber\\
\nonumber
&\leq&C_1\ep\|T_0\|_\infty \left(\int_{|x|<M}\IE |\vep|^2\,dx\right)^{1/2}\times\\
&&\left(\int_{|x|<M}\IE[|\veptil|^2]\IE\lt[|\n^2\veptil|^2\rt]\,dx+
\int_{|x|<M}\IE\lt[|\n\veptil|^4\rt]\,dx\rt)^{1/2}
\nonumber\\
&\leq &C_2\ep\left(\int_{|x|<M}\IE\lt[|\n^2\veptil|^2\rt]\,dx\right)^{1/2}\label{44}
\eeq
by the Gaussianity of the fields. The right side of  (\ref{39}) and (\ref{44})
tends to zero if either
\[
\alpha+2\beta>3
\]
or
\be
\label{47}
\alpha+2\beta=3,\quad\lim_{\ep\to 0}\ep
\sqrt{\log{(1/\ell_1)}}=0
\ee
or
\be
\label{48}
\alpha+2\beta<3,\quad
\lim_{\ep\to 0}\ep\ell_1^{\alpha+2\beta-3}=0
\ee
is satisfied. The term involving $\ep\lan\tvas,\cv(G_\theta^{(2)}\tvas)\ran$ can be
similarly estimated.

The most severe term involving the prefactor $\kappa$ occurs in
$R^\ep_3$ and can be bounded as
\beq
\nonumber
\kappa\ep^2\IE\lt|\lan\tvas,\Delta\cvtil(\cvtil(\theta))\ran\rt|
&\leq& C\kappa\ep^2\|T_0\|_\infty \lt(\int_{|x|<M}\IE\lt[|\n^3\veptil|^2\rt]\rt)^{1/2}\\
&\sim&\left\{\begin{array}{ll}
\kappa\ep^2,&\hbox{for}\,\,\alpha+2\beta>4\\
\kappa\ep^2\sqrt{\log{(1/\ell_1)}},&\hbox{for}\,\,\alpha+2\beta=4\\
\kappa\ep^2\ell_1^{\alpha+2\beta-4},&\hbox{for}\,\,\alpha+2\beta<4
\end{array}\right.
\eeq
the right side of which tends to zero if either
\[
\alpha+2\beta>4
\]
or 
\[
\alpha+2\beta=4,\quad\lim_{\ep\to 0}
\kappa\ep^2\sqrt{\log{(1/\ell_1)}}=0
\]
or
\be
\label{51}
3<\alpha+2\beta<4,\quad\lim_{\ep\to 0}
\kappa\ep^2\ell_1^{\alpha+2\beta-4}=0
\ee
or
\[
2<\alpha+2\beta<3,\quad \lim_{\ep\to 0}
\kappa\ep^2\ell_1^{\alpha+2\beta-4}=
\lim_{\ep\to 0}
\ep\ell_1^{\alpha+2\beta-3}=0.
\]
Note that for $\alpha+2\beta\leq 2$
the condition (\ref{47}) or (\ref{48}) implies
that 
\[
\lim_{\ep\to 0} \ep^2\ell_1^{\alpha+2\beta-4} =0.
\]

Finally we note that in the limit ($\ep,\ell_1\to 0$)
the limiting martingale is, instead of 
(\ref{37}),
\be
M_t(\theta)=f(t)-\int_0^t
\bigg\{f'(s)\left[\frac{\kappa_0}{2}\lan
T_s,\Delta\theta\ran+\frac{1}{a}\bar{A}_3^{(1)}(T_s)\right]
 + \frac{1}{a}f''(s)\bar{A}_2^{(1)}(T_s)\bigg\}\,ds
 \label{37'}
 \ee
 where
 \[
 \bar{A}_2^{(1)}=\lim_{\ell_1\to 0}{A}_2^{(1)},\quad
  \bar{A}_3^{(1)}=\lim_{\ell_1\to 0}{A}_3^{(1)}
  \]
 and 
the limiting process $T_t$
is the (assumed unique) distributional solution to the martingale
problem associated with
the SDE
\beqn
dT_t&=&
\left(\frac{\kappa_0}{2}\Delta +
\frac{1}{a}\bar{\cB}\right)
T_t\,dt+\sqrt{2}a^{-1/2}\n T_t\cdot d\bar{W}^{(1)}_t\\
&=&
\frac{\kappa_0}{2}\Delta
T_t\,dt+\sqrt{2} a^{-1/2}\n T_t \circ d\bar{W}_t^{(1)}
\eeqn
where $\bar{W}_t^{(1)}$ is the Brownian vector field with the spatial
covariance $\bar{\Gamma}^{(1)}(x-y)$ given in (\ref{55})
and the operator $\bar{\cB}$
is given in (\ref{56}).

\commentout{

\subsection{Case 2. $\kappa=0$}
With $\ell_1=\ell_1(\ep)>0$  the stochastic flow $\Phi^{t,\ep}_s(x)$
(\ref{back-flow}) is almost surely well-defined
for $\kappa= 0$. In case of $\kappa=0$, $\Phi^{t,\ep}_s(x)$ is a 
Lebesgue-measure preserving diffeomorphism
for each $t, s,\ep>0$ and we use the representation,
instead of 
\[
\tvas(x)=T_0(\Phi^{t,\ep}_0(x))
\]
in view of which
Proposition~1 clearly holds.  

Here we briefly comment on the necessary modifications of
the proofs of Propositions 2, 3, 4 and 5.
First we note again that all the terms containing $\n\cdot\veptil$
vanish because of incompressibility and all the terms
containing $\kappa$ disappear. 
The proofs of Propositions 2, 3 and 4 are thus unchanged.
Proposition~5 warrants more explanation. 
}

\end{document}